\documentclass[%
reprint,
superscriptaddress,
 amsmath,amssymb,
 aps,
prb,
]{revtex4-2}

\usepackage{lipsum} 

\usepackage{graphicx}
\usepackage{dcolumn}
\usepackage{bm,color}
\usepackage{hyperref}

\newcommand{\dd}{\mathrm{d}}
\newcommand{\Abs}[1]{\left|#1\right|}
\newcommand{\sigmaP}{{\sigma^\prime}}
\newcommand{\Vertex}[3]{{\varGamma^{(#1)}}^{#2}_{#3}}
\newcommand{\ExpValText}[1]{\langle #1 \rangle}
\newcommand{\zg}[1]{{\color{black} #1 }}

\newcommand{\be}{\begin{equation} }
\newcommand{\ee}{\end{equation} }

\begin{document}

\title{Kondo Compensation in a Pseudogap Phase: a Renormalization Group Study}

\author{Csan\'ad Hajd\'u}
 \affiliation{Department of Theoretical Physics, Institute of Physics,
 Budapest University of Technology and Economics, M\"uegyetem rkp. 3., H-1111 Budapest, Hungary}
 \author{C\u at\u alin Pa\c scu Moca}
 \affiliation{HUN-REN—BME Quantum Dynamics and Correlations Research Group,
 Budapest University of Technology and Economics, M\"uegyetem rkp. 3., H-1111 Budapest, Hungary}
 \affiliation{Department of Physics, University of Oradea, 410087, Oradea, Romania}
 \author{Bal\' azs D\'ora}
 \affiliation{Department of Theoretical Physics, Institute of Physics,
 Budapest University of Technology and Economics, M\"uegyetem rkp. 3., H-1111 Budapest, Hungary}
 \author{Ireneusz Weymann}
 \affiliation{Institute of Spintronics and Quantum Information,
 Faculty of Physics and Astronomy, A. Mickiewicz University, 61-614 Pozna\' n, Poland}
 \author{Gergely Zar\'and}
 \affiliation{Department of Theoretical Physics, Institute of Physics,
 Budapest University of Technology and Economics, M\"uegyetem rkp. 3., H-1111 Budapest, Hungary}
 \affiliation{HUN-REN—BME Quantum Dynamics and Correlations Research Group,
 Budapest University of Technology and Economics, M\"uegyetem rkp. 3., H-1111 Budapest, Hungary}

\date{\today}


\begin{abstract}
We investigate the critical behavior of the Kondo compensation in the presence of a power-law pseudogap  in the density of states, 
$\varrho(\omega)\sim |\omega|^\epsilon$. 
For $\epsilon<1$, \zg{generically – in the absence of particle-hole symmetry –} this model exhibits  a quantum phase transition from a partially screened doublet ground 
state to a fully screened many-body 
ground state \zg{upon} increasing \zg {the exchange}  coupling, $j$. 
At the critical  point, $j_c$, the Kondo compensation is found to scale as  
$\kappa(j<j_c) = 1- g(j)$ with the local $g$-factor vanishing as $g \sim |j-j_c|^\beta$. 
We combine perturbative \emph{drone fermion} method with non-perturbative NRG computations to determine the critical 
exponent $\beta (\epsilon)$, which exhibits a non-monotonous behavior as a function of $\epsilon$.  
Our results confirm that the Kondo cloud builds up \emph{continuously} in the presence of a weak pseudogap 
as one approaches the phase transition. 
\end{abstract}
\maketitle

\section{Introduction}

The  Kondo effect~\cite{kondo1964resistance,abrikosov1970theory,nozieres1980kondo,goldhaber1998kondo, hewson1997kondo} 
is one of the most intriguing phenomena in condensed matter physics. As observed 
 by Wilson~\cite{Wilson.1975}, just a single spin coupled antiferromegnetically 
 to a bath of conduction electrons displays the phenomenon of dynamical confinement, whereby
the exchange coupling between the spin and the electrons becomes infinitely large below the 
so-called Kondo temperature, $T_K$. Below this scale,
the conduction electrons screen the magnetic moment of the impurity
by forming the so-called Kondo compensation cloud~\cite{sorensen1996scaling, simon2003kondo, borda2007kondo,bergmann2008quantitative, affleck2010kondo,nuss2015nonequilibrium, borzenets2020observation}.
In the ground state $|G\rangle $, the conduction electrons' total spin, $\mathbf{S}_e $ 
forms a singlet with the impurity spin  $\mathbf{S}$, yielding 
$ \langle G|\mathbf{S}\cdot \mathbf{S}_e |G\rangle = -3/4$. 

\begin{figure}[t]
    \centering
    \includegraphics[width=0.9\columnwidth]{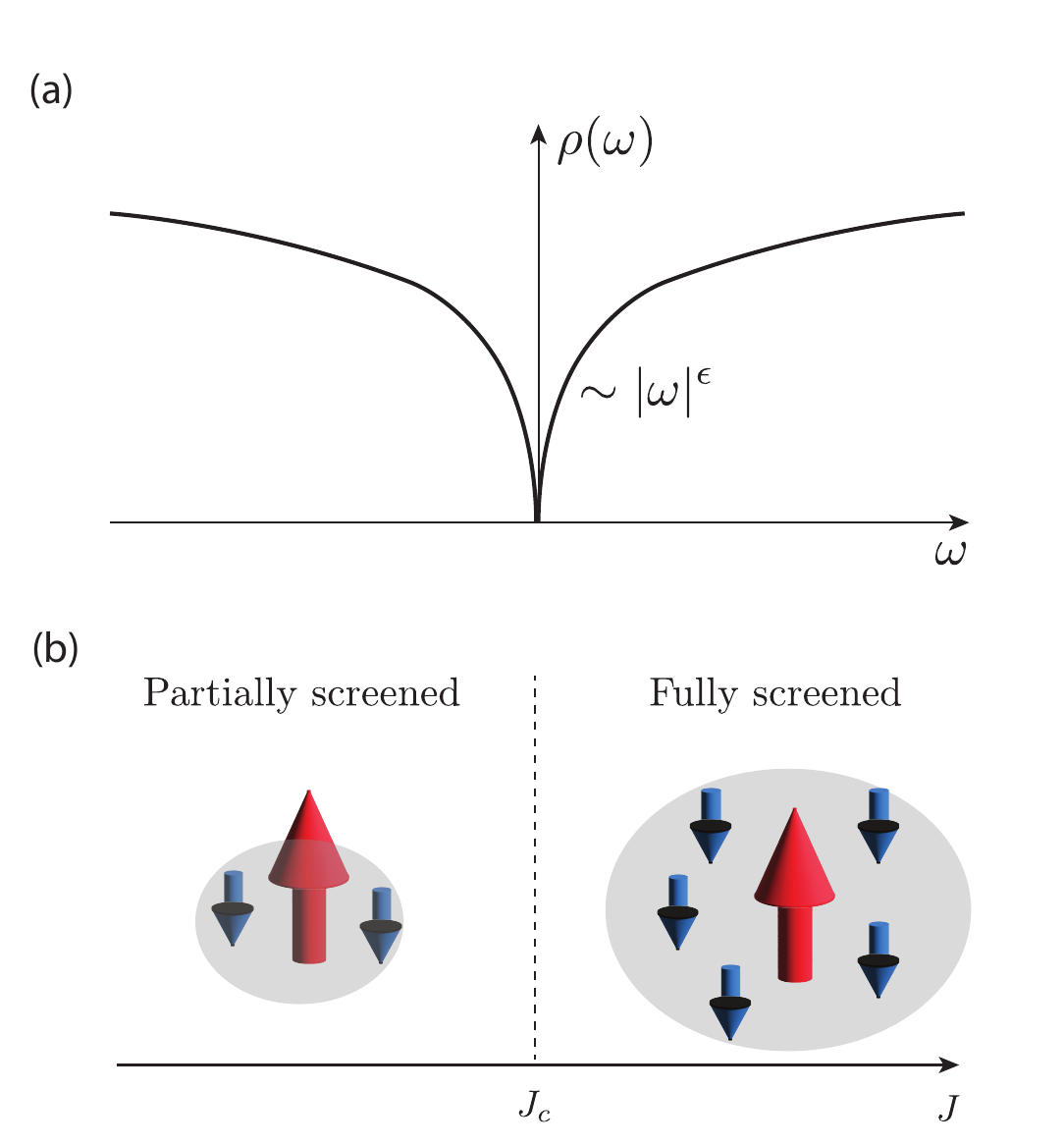}
    \caption{(a) Illustration of a typical pseudogap density of states.
(b) Schematic phase diagram of the \zg{generic, particle-hole asymmetric} model for $0<\epsilon<\zg{1}$, exhibiting a quantum phase transition at a finite Kondo coupling $J_c$. When $J<J_c$, the magnetic impurity is partially screened, whereas in the regime $J>J_c$, the impurity spin is fully compensated.}
    \label{fig:sketch}
\end{figure}

It is an exciting question if confinement and screening persist in a superconductor or in pseudogap systems with a vanishing density of states at the Fermi energy, $\varrho(\omega)\sim |\omega|^\epsilon$, as illustrated in Fig.~\ref{fig:sketch}(a). 
In both cases,  \zg{for a generic, particle-hole asymmetrical system,} a quantum phase transition takes place between the so-called 'doublet' or 'partially screened' phase
and a 'singlet' or 'fully screened' phase upon increasing the exchange coupling $J$ between the impurity and the 
electrons~\cite{Gonzalez.1996,Ingersent.1996, Gonzalez.1998, Simon.1999, Hur.2000,Vojta.2001, Polkovnikov.2001, Matsumoto.2001b, Ingersent.2002,Polkovnikov.2002, Fritz.2004,Lee.2005}, \zg{as} illustrated in Fig.~\ref{fig:sketch}(b).
\zg{In the  special case of particle-hole symmetry, discussed later, the transition is limited to 
$\epsilon <1/2$, and is of somewhat peculiar nature.}

In a superconductor, this transition is referred to as the Yu-Shiba-Rusinov transition~\cite{luh1965bound,shiba1968classical,rusinov1969superconductivity}.
In conventional $s$-wave superconductors, this is a first-order phase transition~\cite{jellinggaard2016tuning, hatter2017scaling, heinrich2018single}, while in pseudogap systems 
the  transition is of first order for a 'strong' pseudogap, $\epsilon >1$, \zg{while it} is of second order in the case of a 'soft' 
pseudogap, $\epsilon <1$. As expected, the transition for $\epsilon =1$ – the case relevant for a $d$-wave superconductor –
is of Kosterlitz-Thouless type~\cite{Polkovnikov.2001, Ingersent.2002, Vojta.2002, Fritz.2004, Lee.2005, cortes2021observation}.  

As shown in Ref.~\cite{Moca.2021}, somewhat counterintuitively, 
even in the 'unscreened' phase,  quantum fluctuations \emph{do \zg{partially} screen} the impurity spin 
 and give rise to a \emph{reduced compensation}. This can be characterized by 
 the compensation strength $\kappa <1$, defined through 
\begin{equation}
\kappa \equiv - \frac 4 3 \langle \mathbf{S}\cdot \mathbf{S}_e \rangle \;.
    \label{eq:sum_rule}    
\end{equation}
For a spin $S=1/2$,  \zg{in the partically screened phase} the compensation $\kappa$ is directly related to the local 
$g$-factor as~\cite{Moca.2021}
\begin{align}
    \kappa = 1-g, 
    &&\quad g = 2 \langle S^z\rangle_{J,h\to 0} ,
    \label{eq:compensation}
\end{align}
with $h$ a local magnetic field acting on $\bold{S}$. 
In a usual s-wave superconductor with a gap $\Delta$, 
the local $g$-factor as well as $\kappa$ are universal functions of the ratio $T_K/\Delta$, and display a universal jump at a critical ratio 
 $(T_K/\Delta)_c$~\cite{Moca.2021}, where the YSR transition takes place.

This picture needs be modified  
for the \zg{pseudogap} Kondo impurity problem, where the conduction electrons exhibit a power-law density of states~\cite{Withoff.1990,Fradkin.1996,Fradkin.1997,Ingersent.1996,Gonzalez.1996,Gonzalez.1998,Simon.1999,Hur.2000,Zhu.2000,Polkovnikov.2001,Vojta.2001, Matsumoto.2001,Matsumoto.2001b,Polkovnikov.2002,Ingersent.2002,Vojta.2002, Dai.2003, Fritz.2004,Fritz.2005, Kircan.2008,Vojta.2010,Fritz.2013},   illustrated 
in Fig. \ref{fig:sketch}(a),
\begin{equation}
        \rho(\omega) =  \rho_0 \Abs{\omega / D}^\epsilon, \quad \omega\in [-D,D],
        \label{eq:DOS}
\end{equation}
with $\rho_0 = (\epsilon+1)/2D$~\footnote{This definition of $\rho_0$ ensures the correct normalization of the density of states. However, in the NRG calculations, we set $\rho_0 = 1/2D$}. 
In the pseudogap Kondo model,   a magnetic impurity of spin $S=1/2$ is coupled to this electron bath through an antiferromagnetic 
exchange Hamiltonian, 
\begin{equation}
    H_{\text{K}} = \frac{1}{2} J \,\sum_{\sigma,\sigma'} \mathbf{S}\cdot\psi^\dagger_\sigma(0)\boldsymbol{\sigma}_{\sigma\sigma'}\psi_{\sigma'}(0).
    \label{eq:Kondo}
\end{equation}
Here, $\bold{s} = \frac{1}{2} \sum_{\sigma,\sigma'} \psi^\dagger_\sigma(0)\boldsymbol{\sigma}_{\sigma\sigma'}\psi_{\sigma'}(0)$ represents the spin density of the conduction electrons at the impurity site (located at the origin), with  
\begin{equation}
    \psi_\sigma(0) = \int_{-D}^{D} \sqrt{\rho(\xi)}\;c_\sigma(\xi)d\xi,
\end{equation}
 denoting the annihilation operator of an electron with spin $\sigma$ at position $x$, and $J$  the (unrenormalized) exchange interaction. 
Here $c_\sigma(\xi)$ is a fermionic operator that annihilates a quasiparticle with energy $\xi$ and spin $\sigma$ and  satisfies the anti-\zg{commutation} relations $\{c_\sigma(\xi), c_{\sigma'}^\dagger(\xi')\} = \delta_{\sigma\sigma'}\delta(\xi-\xi')$.

The dimensionless exchange interaction is defined as $j = \rho_0 J$. 
To obtain the $g$-factor, we apply a small local magnetic field  $h$ along the $z$-direction, as described by the 
Zeeman term, 
\begin{equation}
H_{\text{Z}} = h \, S_z\;.
\end{equation}
Furthermore, we allow for particle-hole symmetry breaking by adding a term
\begin{equation}
    H_{\rm V} = V\;\zg{\sum_\sigma}\;\psi^\dagger_\sigma(0)\psi_{\sigma}(0),
\end{equation}
to the Hamiltonian, with $V$ the strength of the scattering potential. Analogously to the 
exchange interaction, we define the dimensionless scattering strength as  $v = \rho_0 V$.
\begin{figure}[t]
    \centering
    \includegraphics[width=0.7\columnwidth]{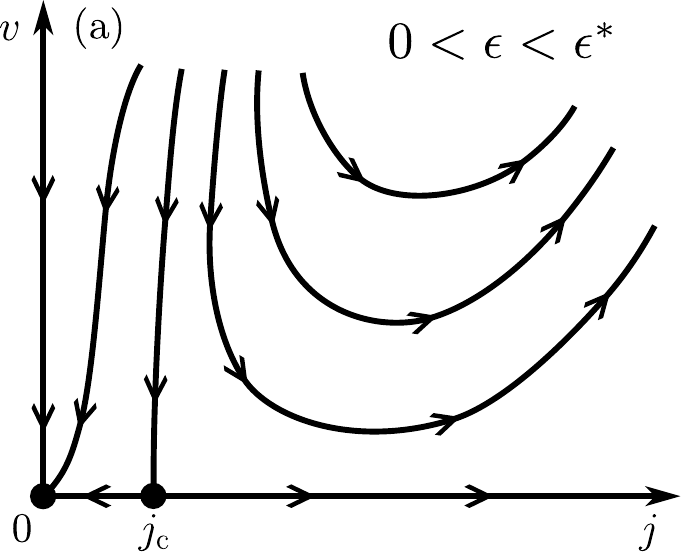}
    \includegraphics[width=0.7\columnwidth]{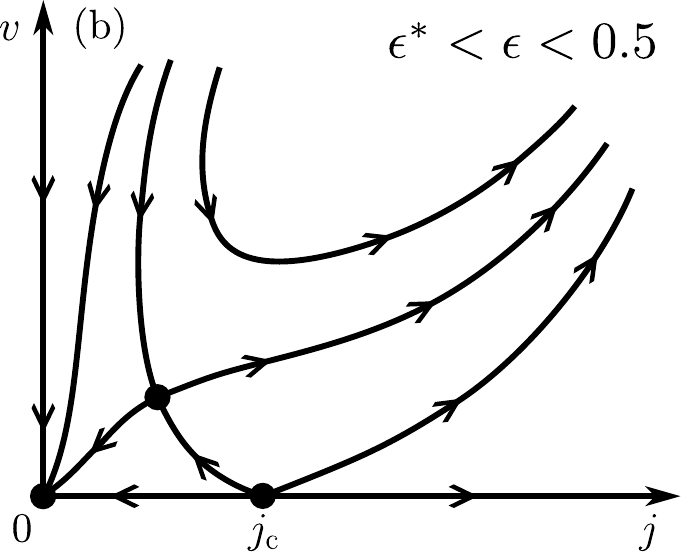}
    \caption{Renormalization group flow diagrams in  $(j,v)$ parameter space for different values of $\epsilon$.
    	(a) When $0<\epsilon <\epsilon^*\approx 0.375$, the particle-hole asymmetry is irrelevant for $j<j_c$.
    	(b) When $\epsilon^*<\epsilon <0.5$, an asymmetric fixed point develops 
    from the symmetric one, whose position changes and moves further away with increasing $\epsilon$.}
    \label{fig:flow_diagrams}
\end{figure}

The phase diagram of the pseudogap Kondo model, 
as first \zg{revealed} by Gonzalez-Buxton and Ingersent in Ref.~\cite{Gonzalez.1998},  
exhibits  greater complexity than \zg{that of} the original Kondo model, corresponding to $\epsilon=0$
(see Fig.~\ref{fig:flow_diagrams}). 
The doublet-singlet phase transition turns out to be of \emph{second order} for all 
$0<\epsilon <1$.  However, particle-hole asymmetry breaking plays a delicate role in this case: 
the particle-hole symmetry-breaking  local scattering $V$
becomes relevant beyond a critical value, $\epsilon>\epsilon^*\approx 0.375$, 
and although the phase transition remains of second order, a novel (still universal) fixed point 
governs the phase transition in the absence of particle-hole symmetry. 
\zg{Furthermore, $V$ is  relevant in the strong coupling phase for any $\epsilon$. The generic behavior therefore 
corresponds to $V\ne0$, and particle-hole symmetry breaking terms cannot be thrown away, 
as usual. Nevertheess, $V$ remains irrelevant in the 'unscreened' phase  for  $\epsilon< 1$ 
as well at the critical point for  $\epsilon<\epsilon^*$   \cite{Gonzalez.1998,Fritz.2004}. }

In this work, we  focus on the fate of \emph{compensation} cloud, and study 
the \emph{critical behavior} of the compensation parameter $\kappa$
across the quantum phase transition~\cite{Ingersent.2002}
by using a combination of analytical and numerical methods.
We use an $\epsilon$-expansion approach by employing  a \emph{Dirac}, or \emph{drone fermion} 
representation~\cite{shnirman2003spin,shchadilova2020fermionic}  of the spin 
to describe the small $\epsilon$ regime within a Multiplicative Renormalization Group (MRG) framework. 
In this approach the spin is represented in terms of an ordinary and a Majorana fermion. 
The drone fermion MRG approach allows us to introduce a small, local magnetic field $h$, 
and extract the corresponding local $g$-factor.
Close to the transition,
the local moment's amplitude is found to display critical behavior,
\begin{equation} 
    g \propto 
    \begin{cases}
        (j_\textrm{c} - j)^\beta, & \quad  j < j_\textrm{c}\ \\ 
        0  ,                  & \quad  j > j_\textrm{c}\; ,
    \end{cases}
    \label{eq:g-factor-scaling}
\end{equation}
signaling a second order phase transition~\cite{Gonzalez.1998,Ingersent.2002}. The MRG approach yields an exponent 
\be 
\beta_\text{MRG} =  \frac  {\epsilon} {  2 \epsilon - 1 + \sqrt{1 - 2 \epsilon}} \,\zg{-1 = \frac \epsilon 2 + \dots \;,}
\label{eq:critical}
\ee 
 in excellent agreement with our numerical renormalization group (NRG)
calculations~\cite{Wilson.1975,legeza2008manual}.

The MRG approach  captures only the electron-hole symmetrical critical behavior for small $\epsilon$. 
However,  as discussed above, the electron-hole symmetrical  
fixed point becomes unstable beyond  $\epsilon^*$, and ceases to exist 
above $\epsilon > 1/2$. Here, a novel fixed point governs the critical behavior in the absence of electron-hole 
symmetry, which lies beyond the 
range of validity  of  the MRG approach. This regime can be captured  
by NRG computations, which  confirm that Eq.~\eqref{eq:g-factor-scaling} remains valid 
even for $\epsilon^*<\epsilon< 1$, but $\beta(\epsilon)$ follows a curve clearly distinct from 
\eqref{eq:critical}. 

\section{The multiplicative renormalization group approach }\label{sec:MRG}

In the context of the Kondo problem, various \zg{spin} representations  exist,
including Abrikosov pseudofermions~\cite{abrikosov2012methods},
Majorana fermions~\cite{mao2003spin}, and the \zg{slave boson} approach~\cite{simon2003kondo}.
In this work, we employ an alternative representation known as the \emph{drone fermion} representation~\cite{shnirman2003spin, shchadilova2020fermionic},
which can be viewed as an extension of the Majorana fermion representation.
In this representation, each spin component is expressed in terms of Majorana fermions,
\begin{align}
S^a = -\frac{i}{4} \epsilon^{abc}\eta^b\eta^c, && {a,b,c}\in{x,y,z},
\end{align}
satisfying the usual anticommutation relations $\{\eta^a,\eta^b \}=2\delta^{ab}$.
However, the presence of the external magnetic field $h$  makes the
Majorana fermion representation challenging to work with. This difficulty can be
overcome by combining the $\eta^{x,y}$ operators to  a Dirac (drone) fermion,
\begin{align}
c^\dagger = \frac{1}{2} (\eta^x +i \eta^y), && c= \frac{1}{2}(\eta^x-i\eta^y), 
\end{align}
which can be used to represent the \zg{spin operators as }
\be
S^+  =  \eta^z\, c^\dagger \;,\quad S^-  =   c\, \eta^z \;, \quad S^z  = c^\dagger c - 1/2 \;. 
\ee
To properly define the associated Hilbert space for the Majorana fermions,
an even number of Majorana fermions is required, necessitating the introduction of a fourth Majorana  particle, $\eta^0$.
Compared to the original Hilbert space 
of the impurity spin $S=1/2$,
which \zg{is of} dimension \zg{two}, 
the new Hilbert space of the Majoranas $(\eta^z, \eta^0)$ and the Dirac fermion $(c,c^\dagger)$ has  four dimensions.
However, \zg{the latter} can be factorized into even and odd parity sectors,
each of dimension $2$. 
As the interaction Hamiltonian
contains only combinations of an even number of particle operators,
the (Majorana)  parity of the spin states remains unchanged,
implying that the two sectors evolve independently and remain completely separated.
In this representation, the Zeeman term can be expressed in terms of the Dirac fermion as
$H_{\text{Z}} = h(c^\dagger c-1/2)$,
while the interaction part (considering the anisotropic exchange interaction) is given by
\begin{equation}
H_{\text{K}} = \psi^\dagger_\sigma(0)\Big(\frac{ J_\perp}2 \big( \eta\, c^\dagger \sigma^-_{\sigma\sigma'} + \text{h.c.}) +  \frac{J_z} 2  c^\dagger c\, \sigma^z_{\sigma\sigma'} \Big) \psi_{\sigma'}(0),
\label{eq:H_K}
\end{equation}
where we introduced the notation $\eta^z\to \eta$, and an implicit summation over spin labels is implied.
Here, we broke the $SU(2)$ symmetry down to the $U(1)$ in the couplings, although we shall focus later on 
the isotropical limit $J_\perp = J^z = J$. 

\begin{figure}
    \centering
    \includegraphics[width=\linewidth]{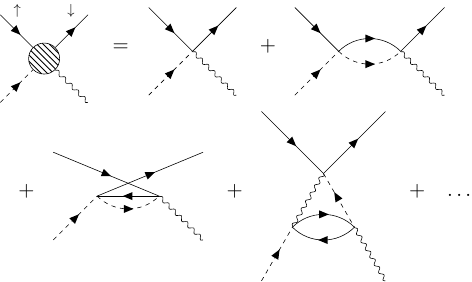}
    \caption{First and second order vertex corrections for $\gamma_{\sigma\sigmaP}^{-}$. 
    The solid line corresponds to the 
    fermionic propagator, dashed line represents the Dirac fermion, while the wavy line 
    represents the Majorana fermion.}
    \label{fig:vertex_corrections}
\end{figure}

For the Kondo problem, the perturbative renormalization group approach  
allows  us to describe the physical properties and capture various phases.
In this framework, an energy cutoff scale, often denoted by $\zg{D}$ is introduced. 
A renormalization step consists in: $(i)$ using perturbation theory to calculate the 
vertex corrections (see Fig.~\ref{fig:vertex_corrections} for typical 
vertex correction diagrams) and the self-energy correction for the dressed Majorana and 
Dirac fermion propagators, 
$F_\eta(t-t')\equiv -i\langle T\eta(t) \eta(t') \rangle$ and 
 $F_c(t-t')\equiv -i\langle T c(t) c^\dagger(t') \rangle$, respectively.
$(ii)$ reducing 
the cut-off $\zg{D} \to \zg{\tilde D} =\zg{D} \,e^{-l}$ by integrating out the energy shell $\textmd{d}\zg{D \approx}-\zg{D} \,\textmd{d}l$,
followed by $(iii)$ a rescaling of the couplings and deriving the renormalization group equations. 

In terms of Eq.~\eqref{eq:H_K}, it is appropriate to define the non-interacting 
vertices as
\begin{align}
    \Vertex{0}{\pm}{\sigma \sigmaP} = \frac{J_\perp}{2} \sigma^\pm_{\sigma \sigmaP}, 
    && \Vertex{0}{z}{\sigma \sigmaP} = \frac{J_z}{2} \sigma^z_{\sigma \sigmaP}.
\end{align}
Due to higher-oder scattering processes, these vertices become renormalized, $ \Gamma^{(0)}\to \Gamma $. 
Similar to the dimensionless coupling $j\equiv \varrho_0 J$, we then
define the dimensionless (dressed) vertices for our model  as  
\begin{align}
    \gamma_{\sigma\sigmaP}^{z}\equiv \rho_0\, \varGamma^{z}_{\sigma\sigmaP}\;,
    && 
    \gamma_{\sigma\sigmaP}^{\pm}\equiv \rho_0\, \varGamma^{\pm}_{\sigma\sigmaP}.
\end{align}
Notice that these are functions of the external frequencies as well as of  the   couplings, 
\zg{$ j_\perp$ and $ j_z$, and the  cut-off, ${ D}$. }

\begin{figure}[t]
    \centering
    \includegraphics[width=0.9\columnwidth]{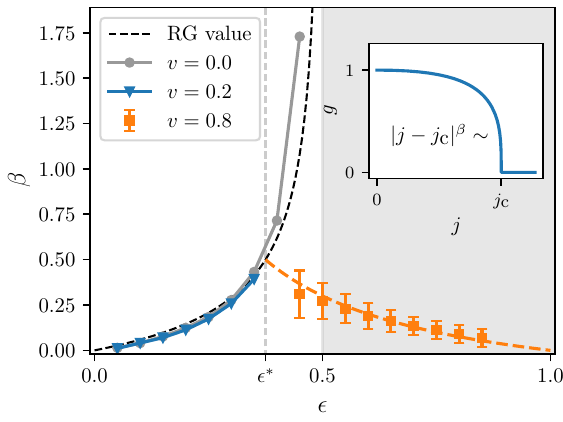}
    \caption{The evolution of the critical exponent $\beta$
   	as a function of the pseudogap exponent $\epsilon$ evaluated using different approaches.
    The dashed line presents the analytical result obtained for the particle-hole \zg{symmetrical} case using the MRG approach,
    while the symbols represent the results obtained from the NRG approach.
    The inset depicts the overall scaling behavior for the $g$-factor as described by Eq. \eqref{eq:g-factor-scaling}.}
    \label{fig:beta}
\end{figure}
\begin{figure}[t]
    \centering
    \includegraphics[width=0.8\columnwidth]{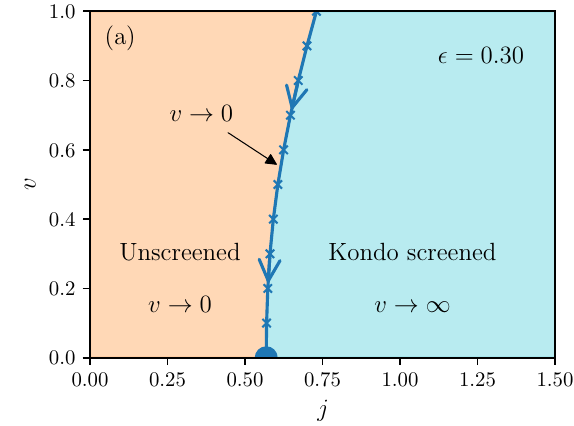}
    \includegraphics[width=0.8\columnwidth]{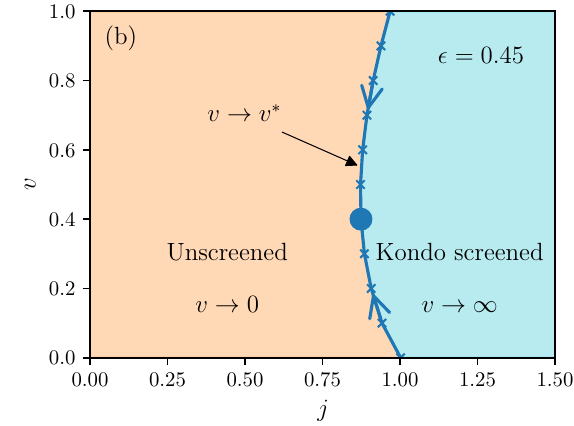}    
    \caption{Phase diagrams extracted from NRG.
    	(a)  Case of  $\epsilon = 0.30 < \epsilon^*$.
    	The renormalization group flow goes to the symmetric fixed point $v = 0$
    	along the critical line.
    	(b) Case of $\epsilon = 0.45 > \epsilon^*$. 
    	The renormalization group flow goes to the asymmetric fixed point, $v = v^*$, along the critical line.
    	The approximate position of this fixed point is shown as a \zg{filled circle}. 
	}
    \label{fig:phase_diagram}
\end{figure}

\begin{figure*}[t]
    \vspace{0.5cm}
    \includegraphics[width=0.6\columnwidth]{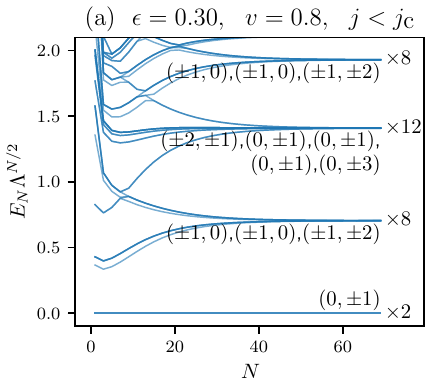}
    \includegraphics[width=0.6\columnwidth]{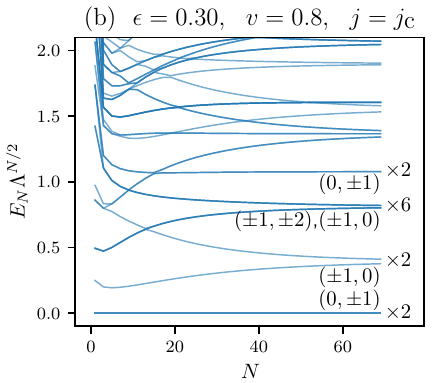}
    \includegraphics[width=0.6\columnwidth]{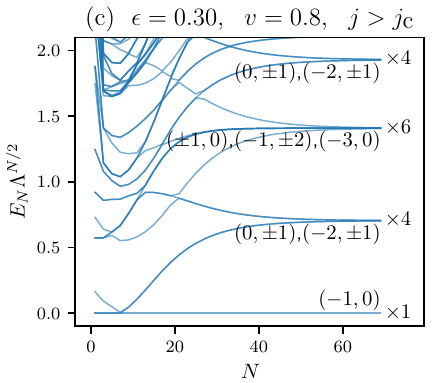}
    \includegraphics[width=0.6\columnwidth]{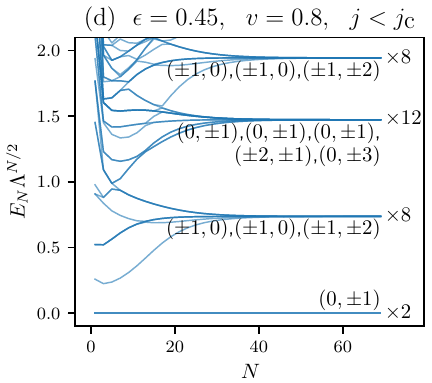}
    \includegraphics[width=0.6\columnwidth]{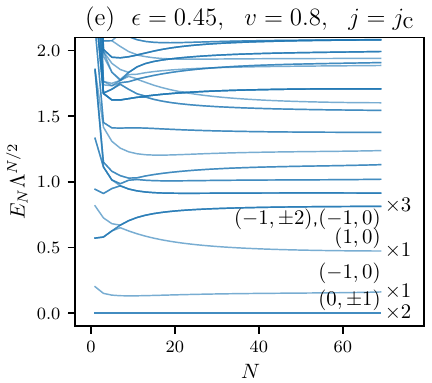}
    \includegraphics[width=0.6\columnwidth]{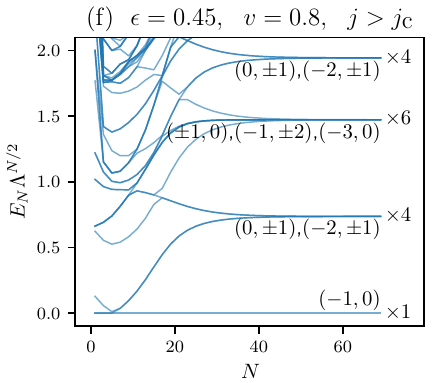}
    
    \caption{NRG finite size spectra are shown for a particular combination of the couplings $V$ and $J$,
    	with two power-law exponents $\epsilon = \{0.30, 0.45\}$.
    	The respective values are specified in the title of each panel.
    	Each energy level is accompanied by the corresponding quantum numbers $(Q, 2 S_z)$
    	representing the total charge and the $z$-component of the spin operator $\bf{S}$,
    	along with the level's degeneracy. \zg{Charge degeneracies indicate the irrelevance of the potential scattering $v$. At the weak coupling fixed point, $j<j_c$, $v$ is always irrelevant (panels (a,d)), while in the strong coupling limit, $j>j_c$, it is always relevant  (panels (c,f)).  At the critical point, $j=j_c$, $v$ is irrelevant for $\epsilon<\epsilon^*$ (panel (b)), while it is relevant for 
$\epsilon>\epsilon^*$ (panel (e)).	}}
     \label{fig:NRG_spectrum}
\end{figure*}

\begin{figure}[t]
    \centering
    \includegraphics[width=0.9\columnwidth]{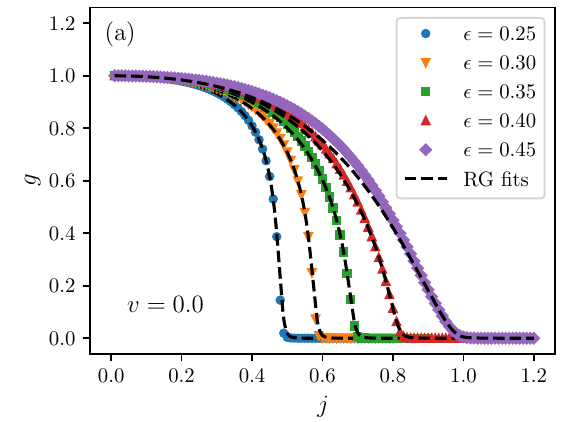}
    \includegraphics[width=0.9\columnwidth]{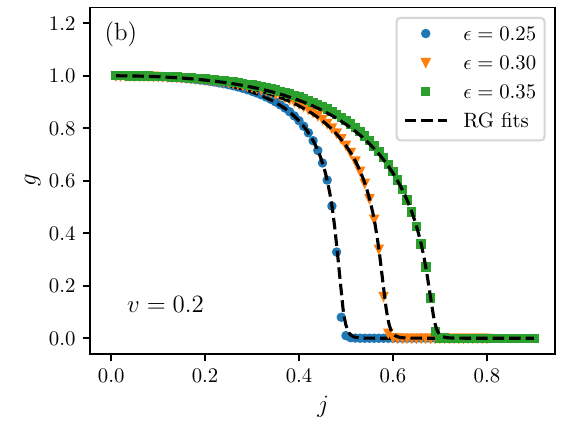}
    \caption{Numerical results for the $g$-factor, computed using NRG,
    	plotted as function of the Kondo coupling $j$.
    	The dashed lines represent the MRG fits.}
    \label{fig:g-factor}
\end{figure}

\begin{figure}[t]
    \centering
    \includegraphics[width=0.9\columnwidth]{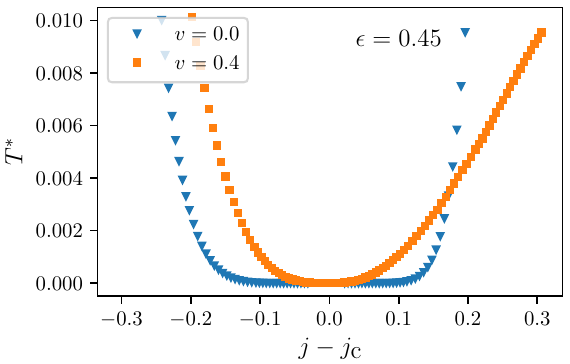}
    \includegraphics[width=0.9\columnwidth]{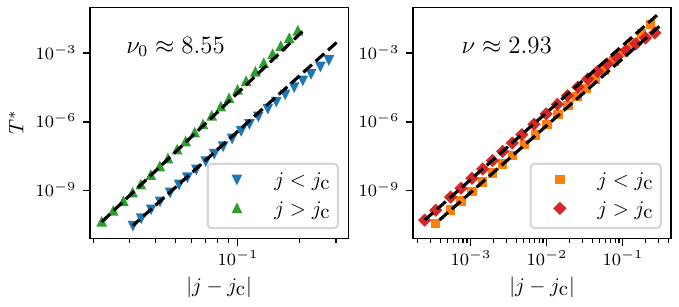}
    \caption{Fermi liquid temperature $T^*$ as extracted from the NRG spectrum for $\epsilon = 0.45 > \epsilon^*$.
    	Quantum critical \zg{behavior is observed  at energies larger than $T^*\sim |j-j_c|^\nu$ with $\nu=1/y_t$.
 	The (unstable) electron-hole 
	symmetrical fixed pont, $v=0$, is characterized by a large exponent, $\nu_{v=0}\approx 8.55$. 
	In the absence of particle-hole symmetry, $v\ne0$, the behavior is close to cubic, $\nu_{v\ne0}\approx 2.93$.}}
    \label{fig:T*1}
\end{figure}

During the renormalization procedure, the dressed Green's functions $F_\eta(\omega)$ and $F_c(\omega)$ do not remain completely invariant,
but acquire  multiplicative  'wave function' renormalization factors, $Z_\zg{\eta}$ and $Z_c$, respectively, 
\zg{$F_\eta(\omega)\to Z_\eta F_\eta(\omega)$ and  $F_c(\omega)\to Z_c F_c(\omega)$.}
Correspondingly, the dimensionless vertex functions  are renormalized as
\begin{align}
    \gamma_{\sigma\sigmaP}^{z}\to Z_c^{-1}\gamma_{\sigma\sigmaP}^{z},
    &&
    \gamma_{\sigma\sigmaP}^{\pm}\to Z_c^{-1/2}Z_{\zg{\eta}}^{-1/2}\gamma_{\sigma\sigmaP}^{\pm}.
    \label{eq:rescaling}
\end{align}
In general, the $Z$-factors are determined order by order in  perturbation theory. 
We   determined  \zg{the vertex corrections up to third order in the couplings, and the self-energy 
corrections up to second order. The $Z$-factors were obtained from the self-energy corrections as,}
\begin{align}
    Z_c = 1 + \frac{j_\perp^2 + j_z^2}{2} \dd l + \dots ~, && Z_\zg{\eta} = 1 + j_\perp^2 \dd l + \dots,
\end{align}
where $\dd l = -\dd \zg{D}/\zg{D}$ represents the infinitesimal change in the dimensionless 
scaling variable $l$. \zg{Eq.~\eqref{eq:rescaling} then} leads to the  scaling equations,
\begin{eqnarray}
    \frac{\dd j_\perp}{\dd l} & = & -\epsilon j_\perp + j_\perp j_z - \frac{j_\perp j_z^2 + j_\perp^3}{4} + \dots  ,\\
    \frac{\dd j_z}{\dd l} & = &-\epsilon j_z + j_\perp^2 - \frac{j_\perp^2 j_z}{2} + \dots ,\\
\frac{\dd h}{\dd l} & = & -{j_\perp^2\over 2} h +\dots .
\end{eqnarray}
In the isotropic case, $j_z = j_\perp$, both equations reduce to the same form, and 
 the scaling equations~\cite{Fritz.2004}  for isotropic coupling $j$
and the external magnetic field $h$ are given by
\begin{eqnarray}
    \frac{\dd j}{\dd l} & = & -\epsilon j + j^2 - \frac{j^3}{2} + \dots ,\\
    \frac{\dd h}{\dd l} & = & -{j^2\over 2} h +\dots .
    \label{eq:h_scaling}
\end{eqnarray}

These admit a trivial "free spin" line of  fixed points  at $j = 0$ and arbitrary $h$, while two non-trivial fixed points exist on the $h = 0$
axis at $j^*_{1,2} = 1 \pm \sqrt{1 - 2 \epsilon}$ for  $\epsilon \leq 0.5$.
From the latter, the  fixed point related to \zg{the screening} phase transition corresponds
to  $j_c \equiv j^*_2 = 1 - \sqrt{1 - 2 \epsilon}= \epsilon + \epsilon^2/2 + \dots$, which is unstable along the  $j$ axis, and therefore describes critical behavior.
Notice that for small $\epsilon$ this fixed point remains close to the origin, $j^*\sim \epsilon$, which enables us to capture the critical behavior perturbatively for small $\epsilon$. \zg{The potential scattering $v$ does not receive vertex corrections, and 
preserves its engineering dimension, $  \dd v /\dd l = -\epsilon\, v $.}

Expanding the renormalization group equations around the critical point
using $\delta j = j - j_\text{c}$ and assuming small $h$, we obtain the following equations
\begin{eqnarray}\label{eq:perturbative}
	\frac{\dd \delta j}{\dd l} & = & ( 2 \epsilon - 1 + 
    \sqrt{1 - 2 \epsilon} ) \delta j + \dots, \\
    \frac{\dd h}{\dd l} & =  &( \epsilon - 1 + \sqrt{1 - 2 \epsilon} ) h + \dots,
\end{eqnarray}
yielding the critical exponents, $y_j = 2 \epsilon - 1 + 
    \sqrt{1 - 2 \epsilon}  = \epsilon-\epsilon^2/2+\dots $ and  $y_h = \epsilon + \sqrt{1 - 2 \epsilon}  = \zg{1}-\epsilon^2/2+\dots$. 
\zg{Notice that, in our formulation, $h$ is a dimensionful quantity, and $y_h $ is the dimension of the dimensionless field, $h/D$, which is relevant
at the fixed point.}

To determine  the compensation $\kappa$, we now compute  local   $g$-factor. 
Due to  the invariance of the free energy under a scale transformation, 
the expectation value of the magnetization satisfies the following scaling relation
\begin{equation} \label{eq:comp:Sz-exp-val}
	\ExpValText{S^z}_{j, \zg{D}} = \Big(\frac{\partial \tilde h}{\partial h}\Big) \ExpValText{S^z}_{\tilde j, \zg{\tilde D}},
\end{equation}
which implies that the $g$-factor can be expressed from the 
scaling equation \eqref{eq:h_scaling} as
\begin{equation} \label{eq:g-factor-integral}
	g = \frac{\partial \tilde h}{\partial h} \;\zg{\to} \;\frac{\zg{\tilde h}}{h} \approx \exp\Big \{-\int\limits_0^\infty \dd l\ \frac{j^2(l)}{2}\Big\}.
\end{equation}
Using \eqref{eq:g-factor-integral} and the scaling equation~\eqref{eq:h_scaling}, this yields the critical exponent
$\beta\zg{={|y_h -1|/y_j = \epsilon/2 + \dots }}$ in Eq.~\eqref{eq:critical}, presented as a dashed line in Fig.~\ref{fig:beta}.
Eq.~\eqref{eq:g-factor-integral} also allows us to 
 evaluate the $g$-factor numerically with some  initial condition $j(0) = j_0$.

\section{Numerical renormalization group approach}

We can extract the behavior of the $g$-factor using Wilson's renormalization group by computing the expectation value 
$g = 2 \langle S^z\rangle_{h\to 0}$.  As shown in Fig.~\ref{fig:beta}, 
the analytical result of Eq.~\eqref{eq:critical} is clearly confirmed 
by our  NRG results  for $\epsilon < \epsilon^*$, where $\beta$ remains unaffected by the breaking of 
electron-hole symmetry, and closely follows the curve predicted by Eq.~\eqref{eq:critical}.
\zg{Strictly speaking, Eq.~\eqref{eq:critical} is valid only to linear order in $\epsilon$, however, the 
result of the next to leading logarithmic scaling equations provides a surprisingly good approximation 
for the numerically obtained exponent, $\beta$.}

Although electron-hole symmetry breaking is  relevant for $\epsilon>\epsilon^*$,
the exponent predicted by  multiplicative RG gives still an excellent estimate  
for $\epsilon^*<\epsilon<1/2$ in the absence of electron-hole symmetry breaking ($v=0)$. 
For $\epsilon>\epsilon^*$ and $v\ne0$, however, we can only determine the critical exponent $\beta$ numerically. 
Here, NRG confirms that the quantum critical spectrum is universal (see Fig.~\ref{fig:NRG_spectrum}), i.e., 
it does \emph{not} depend on 
the specific value of $v$.  Our numerical data in Fig.~\ref{fig:beta} suggest that  $\beta$ approaches $\beta^*\approx 0.5$ at $\epsilon=\epsilon^*$, 
while it decays towards zero as $\epsilon\to 1$.

Figure~\ref{fig:phase_diagram} presents the numerically computed phase diagram 
together with critical points obtained from the NRG finite size spectra.  
The NRG spectrum serves as a powerful tool, allowing to extract the information
that may not be accessible through other means. It enables us to locate precisely the fixed points,
determine the critical values for the couplings associated with quantum phase transitions in the model, or analyze the relevance/irrelevance of operators.
Typical results for  NRG finite size spectra are presented in Fig.~\ref{fig:NRG_spectrum}.
For every energy level in the spectrum, we provide the corresponding quantum numbers $(Q, 2S_z)$ for the $U(1)$ total charge and spin, along with the level's degeneracy. 
For instance, in the first column of Fig.~\ref{fig:NRG_spectrum}\zg{,} corresponding to $j<j_c$ and $v=0.8$, 
both fixed points for $\epsilon=0.30$ and $\epsilon=0.45$ resemble the one 
corresponding to $v=0$, and states with charges $\pm Q$  are degenerate, indicating
the irrelevance of electron-hole asymmetry in this limit. 
Moreover, the ground state appears as a doublet phase $(Q=0, S_z=\pm 1/2)$
with the impurity spin fully decoupled from the environment, as indicated by its degeneracy.

In contrast, the spectrum at the critical coupling $j=j_c(v)$
significantly differs for $\epsilon=0.30$ and $\epsilon=0.45$,
signaling the distinct nature of the two fixed points.
Indeed, for  $\epsilon=0.30<\epsilon^*$ states in the critical spectrum are charge degenerate, 
indicating that $v$ is irrelevant. In contrast, for  $\epsilon=0.45>\epsilon^*$ charge degeneracy is removed
in the critical spectrum, clearly indicating that \zg{$v\to v^* $ finite} at criticality. 

As we increase the coupling $j$ beyond $j_c$, we cross over to the strongly coupled regime. 
Here the ground state transforms into a singlet $(Q=-1, S_z =0)$,
indicating a quantum phase transition in the model, and charge degeneracies are removed, signaling that 
electron-hole symmetry breaking is relevant in the Kondo-screened phase.

In Fig.~\ref{fig:g-factor} we present 
typical results for the $g$-factor for different pseudogap exponents $\epsilon$ and different scattering potentials $v$, as extracted from NRG, and compared with the prediction of the MRG equations. These numerical results allow us to examine in more detail the quantum phase transition. First, notice that the transition in Fig.~\ref{fig:g-factor} happens continuously, implying a second-order phase transition, while the $g$-factor approaches zero in a power-law fashion, according to Eq.~\eqref{eq:g-factor-scaling}, with some power-law $\beta$ that has been determined numerically and displayed in Fig.~\ref{fig:beta}.

After precisely determining the quantum critical point, let us shift our focus on perturbing the quantum critical state intentionally. 
Perturbations such as the Zeeman splitting \zg{or deviations from the critical coupling} typically induce a characteristic Fermi liquid scale, denoted as $T^*$.  
\zg{For $j\approx j_c$, quantum critical behavior is observed  at energies larger than $T^*\sim |j-j_c|^\nu$ with $\nu=1/y_t$ \cite{Fritz.2004}, 
and $y_t$ the RG eigenvalue associated with the relevant ('thermal') operator at the critical pont.
In the presence of electron-hole symmetry, $v=0$, this is simply the eigenvalue associated with $j$, $y^{v=0}_t=y^{v=0}_j$, 
while for $\epsilon>\epsilon^*$ and $v\ne0$, this is the relevant eigenvalue associated with the emerging electron-hole asymmetrical 
quantum critical point.}

\zg{In Fig.~\ref{fig:T*1}
 we illustrate,  how precise manipulation of the exchange coupling $j$ results in \zg{an emerging}
 $T^*$ scale upon varying  the detuning $j-j_c$ for $\epsilon=0.45>\epsilon^*$.
The Fermi liquid scale  $T^*$  displays a power law  scaling
on both sides of the transition, as shown in the 
lower panels of Fig.~\ref{fig:T*1}.
The $j>j_c$  and $j<j_c$ branches show  similar scaling, but with a different prefactor. The (unstable) electron-hole 
symmetrical fixed pont, $v=0$, is characterized by a remarkably  large exponent, $\nu_{v=0}\approx 8.55$, while the theoretically predicted value is $\nu_{\text{Th}} \approx 4.62$. 
The exponent is substantially reduced  in the absence of particle-hole 
symmetry, and we obtain $\nu_{v\ne0}\approx 2.93$ for  $\epsilon=0.45$. 

We remark that both $\nu_{v\ne0}$ and $\nu_{v=0}$  are functions of  $\epsilon$; 
the  exponent  $\nu_{v=0}$  increases with $\epsilon$,  and diverges at $\epsilon=1/2$, 
where the $v=0$ fixed point ceases to exist.  The two exponents are equal at $\epsilon=\epsilon^*$. 
}

\section{Conclusions and discussions}

In this work, we investigated the Kondo compensation mechanism in a pseudogap phase by employing a combination of perturbative and numerical renormalization group techniques. Specifically, we focused on the critical behavior of the \zg{compensation} as the Kondo coupling, $j$, approaches the critical value $j_c$. \zg{Particle-hole symmetry breaking for this transition plays a crucial role, 
and – unlike the  Kondo problem in a metal, it cannot be neglected~\cite{Ingersent.1996,Gonzalez.1998,Ingersent.2002,Vojta.2001,Vojta.2002,Fritz.2004}, since it impacts the 
screened strong coupling phase. In the generic, particle-hole asymmetrical ($v\ne0$) case we find that 
 the system undergoes a second order quantum phase transition  from a partially screened doublet state to a fully screened singlet state, irrespective of the exponent $\epsilon<1$.} 

\zg{Our analysis reveals that, near the critical coupling, the compensation factor $\kappa$ behaves as $\kappa(j < j_c) = 1 - g(j)$, with the local g-factor vanishing as $g \sim |j - j_c|^\beta$, where $\beta(\epsilon)$ is the critical exponent. Kondo compensation thus builds up \emph{continuously} as one approaches the phase transition, and the impurity spin is partially screened even in the weak coupling, 
'unscreened' phase.
We determined the exponent $\beta$ analytically as a function of  $\epsilon$ for small $\epsilon$, and corroborated 
our findings with numerical renormalization group (NRG) calculations. The exponent $\beta(\epsilon)$ exhibits a non-monotonic dependence on $\epsilon$, with distinct behavior emerging for  $\epsilon<\epsilon^*\approx  0.375$ and 
$\epsilon>\epsilon^*\approx  0.375$.
This non-monotonic behavior is due to the emergence of a novel, particle-hole symmetry breaking critical point 
with $v\to v^*= \text{finite}$, governing the critical behavior for $\epsilon>\epsilon^*$. In this regime, we had to 
rely entirely on numerics while determining the critical behavior.}

\zg{As discussed above, particle-hole symmetry breaking plays an important and 
delicate role, and the particle-hole symmetrical, $v=0$ model is somewhat pathological. 
It is, in particular, worth discussing the strong coupling phase, which is unstable against any 
finite $v\ne0$.  Numerically, we find that even for $v=0$, the local $g$-factor scales continuously to 
zero as $j\to j_c$, and it \emph{remains} zero in the strong coupling phase, $j>j_c$. This implies that the 
impurity spin is \emph{screened}. However, as discussed in Refs.~\cite{Gonzalez.1998,Fritz.2004}, for $v=0$ the strong coupling phase 
has a residual entropy. These observations indicate that, although the spin is locally screened, non-Fermi liquid degrees 
of freedom survive in the electronic sector. The fact that they disappear for any finite $v$ could indicate that 
these degrees of freedom may be associated with charge rather than spin 
degrees of freedom. }

\zg{
Finally, we extracted the Fermi liquid scale $T^*$ from the finite size spectrum and analyzed its behavior 
around the quantum critical point. We verified that $T^*$ exhibits a power 
law dependence~\cite{Fritz.2004},  $T^*\sim |j - j_c|^\nu$, marking the boundary between the quantum critical and Fermi liquid regimes. 
This characteristic scale is crucial in understanding how perturbations, such as a local magnetic field 
influence the behavior of the system near the critical point.}

Reaching the quantum critical regime $\epsilon<1$ in an experimental system is somewhat challenging. 
Three-dimensional systems with  quadratic band touching such as  $\alpha$-Sn, HgTe, and pyrochlore Pr$_2$Ir$_2$O$_7$
\cite{zaheer2013spin, cheng2017dielectric, zhang2022large} realize the case $\epsilon=0.5$. In these systems, Coulomb interaction between the conduction electrons 
may give rise to further complications and non-Fermi liquid physics, as initially suggested by Abrikosov and Beneslavskii~\cite{abrikosov1971some, abrikosov1996possible, janssen2015nematic,janssen2016excitonic, zhang2018engineering, tchoumakov2019dielectric, link2020hydrodynamic}.

\begin{acknowledgments}
We would like to thank Matthias Vojta and Ilya Vekhter for the insightful 
discussions and valuable feedback.
This research is supported by the National Research,
 Development and Innovation Oﬃce - NKFIH within
the Quantum Technology National Excellence Program under
Projects No. 2017-1.2.1-NKP-2017-00001 and K142179. 
C.P.M acknowledges support from CNCS/CCCDI–UEFISCDI, under projects number PN-IV-P1-PCE-2023-0159. 
I.W. acknowledges financial support from the National Science Centre
in Poland through the Project No. 2022/45/B/ST3/02826.
\end{acknowledgments}

\bibliography{apssamp}

\end{document}